\documentclass[aps,prc,twocolumn,superscriptaddress, amsmath,amssymb, nofootinbib, nolongbibliography]{revtex4-2}

\usepackage[english]{babel}
\usepackage[T1]{fontenc}
\usepackage{graphicx}
\usepackage{dcolumn}
\usepackage{booktabs}
\usepackage{amsbsy}
\usepackage{dblfloatfix}
\usepackage{siunitx}
\usepackage[svgnames]{xcolor}

\newcolumntype{d}[1]{D{.}{.}{#1}} 
\usepackage[colorlinks]{hyperref}
\hypersetup{breaklinks=true, linkcolor=Navy, filecolor=Navy, citecolor=Navy, urlcolor=Navy}

\usepackage[alph]{parnotes}
\makeatletter
\def\parnoteclear{%
    \gdef\PN@text{}%
    \parnotereset
}
\makeatother

\begin{document}

\title{$l$-forbidden $\mathbf{M1}$ strengths near $^{100}$Sn from
  knockout reactions in Cd and Sn}
  \author{T.~J.~Gray}
  \affiliation{Department of Physics and Astronomy, University of Tennessee Knoxville, Tennessee 37966, USA}
  \affiliation{Physics Division, Oak Ridge National Laboratory, Oak
    Ridge Tennessee 37831, USA}
\author{K.~L.~Jones}
\affiliation{Department of Physics and Astronomy, University of
  Tennessee Knoxville, Tennessee 37966, USA}
\author{R.~Grzywacz}
\affiliation{Department of Physics and Astronomy, University of
  Tennessee Knoxville, Tennessee 37966, USA}

\author{B.~A.~Brown}
\affiliation{Department of Physics and Astronomy, Michigan State
  University, East Lansing, Michigan, 48824, USA}
\affiliation{Facility for Rare Isotope Beams, Michigan State
  University, East Lansing, Michigan, 48824, USA}
\author{A.~Gade}
\affiliation{Department of Physics and Astronomy, Michigan State
  University, East Lansing, Michigan, 48824, USA}
\affiliation{Facility for Rare Isotope Beams, Michigan State
  University, East Lansing, Michigan, 48824, USA}
\author{B.~C.~He}
\affiliation{Department of Physics and Astronomy, University of Tennessee Knoxville, Tennessee 37966, USA}
\author{T.~Miyagi}
\affiliation{Center for Computational Sciences, University of Tsukuba,
  1-1-1 Tennodai, Tsukuba, Ibraki 305-8577, Japan}
\author{A.~Peter}
\affiliation{Department of Physics and Astronomy, University of Tennessee Knoxville, Tennessee 37966, USA}

\author{M.~J.~Basson}
\affiliation{Department of Physics and Astronomy, Michigan State
  University, East Lansing, Michigan, 48824, USA}
\affiliation{Facility for Rare Isotope Beams, Michigan State
  University, East Lansing, Michigan, 48824, USA}
\author{T.~Beck}
\affiliation{Facility for Rare Isotope Beams, Michigan State
  University, East Lansing, Michigan, 48824, USA}
\altaffiliation[Present address: ]{Instituut voor Kern- en Stralingsfysica, KU Leuven, 3001 Leuven, Belgium}
\author{C.~M.~Campbell}
\affiliation{Nuclear Science Division, Lawrence Berkeley National
  Laboratory, Berkeley, California, 94720, USA}
\author{G.~Cerizza}
\affiliation{Facility for Rare Isotope Beams, Michigan State University, East Lansing, Michigan, 48824, USA}
\author{J.~Chung-Jung}
\affiliation{Department of Physics and Astronomy, Michigan State
  University, East Lansing, Michigan, 48824, USA}
\affiliation{Facility for Rare Isotope Beams, Michigan State
  University, East Lansing, Michigan, 48824, USA}
\author{I.~Cox}
\affiliation{Department of Physics and Astronomy, University of Tennessee Knoxville, Tennessee 37966, USA}
\author{P.~Farris}
\affiliation{Department of Physics and Astronomy, Michigan State
  University, East Lansing, Michigan, 48824, USA}
\affiliation{Facility for Rare Isotope Beams, Michigan State
  University, East Lansing, Michigan, 48824, USA}
\author{R.~Ghimire}
\affiliation{Lawrence Livermore National Laboratory, Livermore,
  California 94550, USA}
\author{S.~Gillespie}
\affiliation{Facility for Rare Isotope Beams, Michigan State University, East Lansing, Michigan, 48824, USA}
\author{M.~Grinder}
\affiliation{Department of Physics and Astronomy, Rutgers University,
  New Brunswick, New Jersey 08903}
\author{A.~Hill}
\affiliation{Department of Physics and Astronomy, Michigan State
  University, East Lansing, Michigan, 48824, USA}
\affiliation{Facility for Rare Isotope Beams, Michigan State
  University, East Lansing, Michigan, 48824, USA}
\author{S.~D.~Pain}
\affiliation{Physics Division, Oak Ridge National Laboratory, Oak
  Ridge Tennessee 37831, USA}
\affiliation{Department of Physics and Astronomy, University of Tennessee Knoxville, Tennessee 37966, USA}
\author{A.~Palmisano-Kyle}
\affiliation{Department of Physics and Astronomy, University of Tennessee Knoxville, Tennessee 37966, USA}
\author{K.~P.~Rykaczewski}
\affiliation{Physics Division, Oak Ridge National Laboratory, Oak
  Ridge Tennessee 37831, USA}
\author{D.~Weisshaar}
\affiliation{Facility for Rare Isotope Beams, Michigan State University, East Lansing, Michigan, 48824, USA}
\author{M.~Williams}
\affiliation{Lawrence Livermore National Laboratory, Livermore, California 94550, USA}

\begin{abstract}
Neutron knockout reactions on beams of $^{104,102}$Cd, and $^{104}$Sn
are presented. States in the residual $^{103,101}$Cd and
$^{103}$Sn nuclei are populated, including low-lying $7/2^+$ states of $\nu g_{7/2}$ character. These states have half-lives
$\approx 400$ ps due to their low energy and hindered $B(M1; 7/2^+
\rightarrow 5/2^+)$ strengths. The excited-state half-lives were measured using their
Doppler-shifted lineshapes, and the resulting $B(M1)$ strengths are
compared to Valence Space In Medium Similarity Renormalization Group
(VS-IMSRG) calculations. The VS-IMSRG calculations under-predict the
$l$-forbidden $M1$ strengths in the $^{100}$Sn region, as well as in
other regions of the nuclear chart near $^{40}$Ca and $^{208}$Pb.
\end{abstract}

\maketitle

\section{Introduction}

%


Nuclear magnetic dipole moments are an important nuclear observable
for understanding the microscopic properties of the atomic nucleus~\cite{Sassarini2022,
  Miyagi2024, Wibowo2025}. A cornerstone of nuclear
structure, the nuclear shell model, makes simple predictions for the
values of the magnetic dipole moment in the limit of pure
single-particle wavefunctions near closed shells~\cite{Schmidt1937,Arima2011}. These simple
estimates are known as the Schmidt values. It is well known that
experimental values are often quenched relative to the Schimdt values, and
explaining or predicting the deviations requires a more complex
description of the nucleus.

The usual approach in the shell-model framework is to introduce an
effective $M1$ operator, where deviations from the ``bare'' operator arise from four principal
contributions. These are (1) core polarization or mixing with
excitations from outside valence space, (2) meson-exchange currents,
(3), isobar currents, involving the excitation of one or more nucleons
to their first-excited state --- the $\Delta(1232)$ isobar state~\cite{Pascalutsa2007,Rodriguez2022}, and finally (4) relativistic
effects. References~\cite{Towner1983,Towner1987} provide theoretical background for
the effictive $M1$ operator, which is usually formulated as

\begin{equation}
  \pmb{\mu}^{(i)} = g_L^{(i)}\mathbf{L} + g_S^{(i)}\mathbf{S} +
  g_P^{(i)}[\mathbf{Y_2} \times \mathbf{S}].
  \label{eq:effective_M1}
\end{equation}
Where $i$ refers to the nucleon charge state ($p$ for proton, $n$ for
neutron) and $\mathbf{Y_2}$ is a spherical tensor. There are three coupling constants: the orbital ($g_L$), spin
($g_S$), and tensor ($g_P$) constants. The ``bare'' operator refers to the values of $g$ for the free proton and neutron: $g_L^p = 1, g_L^n = 0, g_S^p = 5.586, g_S^n = -3.826$, and $g_P^p = g_P^n = 0$. These coupling constants parametrize the
effective operator and can be derived from theory. Accurate
calculations of these coupling constants is critical for predictions of not only magnetic moments, but also $M1$ transition
strengths. Gamow-Teller strengths are also affected, since meson-exchange and isobar
currents play a dominant role in their
quenching~\cite{Towner1983, Towner1987, Gysbers2019}. We draw particular attention
to the $g_P$ coupling term here.

\textit{Ab initio} theory has made much progress in recent years
pushing towards higher mass regions~\cite{Miyagi2022,Stroberg2021,Stroberg2017}. The inclusion of
meson-exchange currents is crucial for applying \textit{ab initio} theories to $M1$ observables~\cite{Sassarini2022, Miyagi2024, Wibowo2025}.

The coupling constants in Eq.~\ref{eq:effective_M1} can also be
constrained experimentally. Brown and Wildenthal performed fits to
empirical data across the $sd$ shell~\cite{Brown1983, Brown1987}. The
data contained magnetic moments, as well as $B(GT)$
strengths and some $B(M1)$ transition strengths. While the magnetic
moment and $B(GT)$ data determines $g_L$ and $g_S$, the tensor term
$g_P$ is predominantly constrained by $l$-forbidden $M1$ transitions
--- in the case of $sd$ shell nuclei between the $0s_{1/2}$ and
$1d_{3/2}$ orbits. Regarding the $g_P$ coupling term, the authors
conclude that the emprical corrrection is almost twice as large as
calculated in Ref.~\cite{Arima2011}. The $sd$ shell
nucleus $^{45}$Cl was also investigated in Ref.~\cite{Stroberg2012}; the $l$-forbidden $M1$ strength is
sensitive to off-diagonal $sd$-$fp$ cross-shell matrix elements. With
a ground-state spin $J^{\pi}_{\mathrm{g.s.}} =
3/2^+$~\cite{Stroberg2012,Bhattacharya2023,Tripathi2024}, the $g_P$ required for $^{45}$Cl is
$\approx 50\%$ larger than rest of the $sd$ shell~\cite{Stroberg2012}.

Beyond the $sd$ shell, the $^{57}$Cu-$^{57}$Ni pair has been
studied in Ref.~\cite{Semon1996}, and once again
the $l$-forbidden $M1$ transition strength between the $0f_{5/2}$ and
$1p_{3/2}$ orbits constrains the $g_P$ coupling term. Similar to Ref.~\cite{Brown1987}, the authors find
a discrepancy between the emprical $g_P$ and that predicted by theory.

The next-heaviest region where $l$-forbidden $M1$ transitions may be
investigated near a double shell closure is the $^{100}$Sn region,
with the $0g_{7/2}$ and $1d_{5/2}$ orbits. $^{100}$Sn is the heaviest doubly-magic $N=Z$ nucleus, and
additionally represents the end of the astrophysical rp-process~\cite{Schatz2001}. Experimental constraints of the
single-particle orbits outside $^{100}$Sn are crucial to understand
the region. Major questions with respect to these single-particle orbits
still remain: namely the ordering of the $0g_{7/2}$
and $1d_{5/2}$ neutron orbitals in $^{101}$Sn remains experimentally
unverified~\cite{Sewerniak2007, Darby2010}.

Here we present neutron knockout experiments into $^{103}$Cd,
$^{101}$Cd, and $^{103}$Sn. We strongly populate low-lying $0g_{7/2}$
excited states, and measure their half-lives. These half-lives are
converted into $B(M1)$ transition strengths, and comapred to
state-of-the-art shell model and \textit{ab initio} calculations,
providing a stringent test of the effective $M1$ operator near
$^{100}$Sn.

$^{103}_{48}$Cd$_{55}$ has been studied with $\beta$ %
decay~\cite{Wouters1983, Verplancke1984, Bom1988, Szerypo1997,
  Karny1998}, heavy-ion reactions~\cite{Meyer1980, Palacz1997, Chakraborty2007}, and fast-timing~\cite{Kisyov2011}. This is the first
instance of a direct reactions study.

Literature on the more exotic $^{101}$Cd is much more sparse, with
only a single decay study~\cite{Huyse1988}, and a single in-beam
heavy-ion reaction~\cite{Alber1992}. Studies on $^{103}$Sn are
likewise sparse, with an $\alpha$-decay
measurement~\cite{Sewernyiak2002} and an in-beam spectroscopy
study~\cite{Fahlander2001} giving the only information on excited
states.

\section{Methods}
An experiment was conducted at the Facility for Rare Isotope Beams (FRIB). A primary beam of 227~MeV/nucleon $^{124}$Xe was fragmented on a $^{12}$C production target. The Advanced Rare Isotope Separator (ARIS) was used to separate the fragments, delivering beams of $^{104,106}$Sn at 108~MeV/nucleon and 82~MeV/nucleon, respectively, to the analysis line of the S800 spectrograph~\cite{Bazin2003}. The Sn beams were incident on a $^{9}$Be target, 47~mg/cm$^2$ thick. The residual nuclei were analysed by the S800 spectrograph in dispersion-matched mode, allowing the unreacted beam species to be blocked and maximising the momentum resolution. The GRETINA array~\cite{GRETINA} with 12 quads (48 crystals) surrounded the target chamber.

\paragraph*{}
The purity of the incoming beam was low ($\approx 1 \%$
and $\approx 8\%$ for $^{104,106}$Sn settings, respectively). The
contamination present was primarily in the form of species with $(Z-i,
A-i)$ relative to the nominal setting: i.e. for $^{106}$Sn, major
contaminants were $^{105}$In, $^{104}$Cd, and $^{103}$Ag. The incoming
beam species can be established on an event-by-event basis with the
TOF-TOF method: comparing the time differences between the plastic
timing scintillator at the focal plane of the S800, and the upstream
plastic timing scintillators at the DB3 position of FRIB's fragment
separator and the S800 analysis line object location, respectively. For analysis details see
Refs.~\cite{Ayres2014, Longfellow2020, McDaniel2011, Stroberg2014}. The present paper focuses on knockout reactions on the contaminant Cd species, $^{104,102}$Cd, as these are the strongest-populated even-even nuclei in the data set. Results on the very low statistics $^9$Be($^{104}$Sn,$^{103}$Sn+$\gamma$)$X$ reaction are also presented here, while a future publication will cover the $^9$Be($^{106}$Sn,$^{105}$Sn+$\gamma$)$X$ reaction.

\begin{figure}[t]
  \includegraphics[width=0.9\columnwidth]{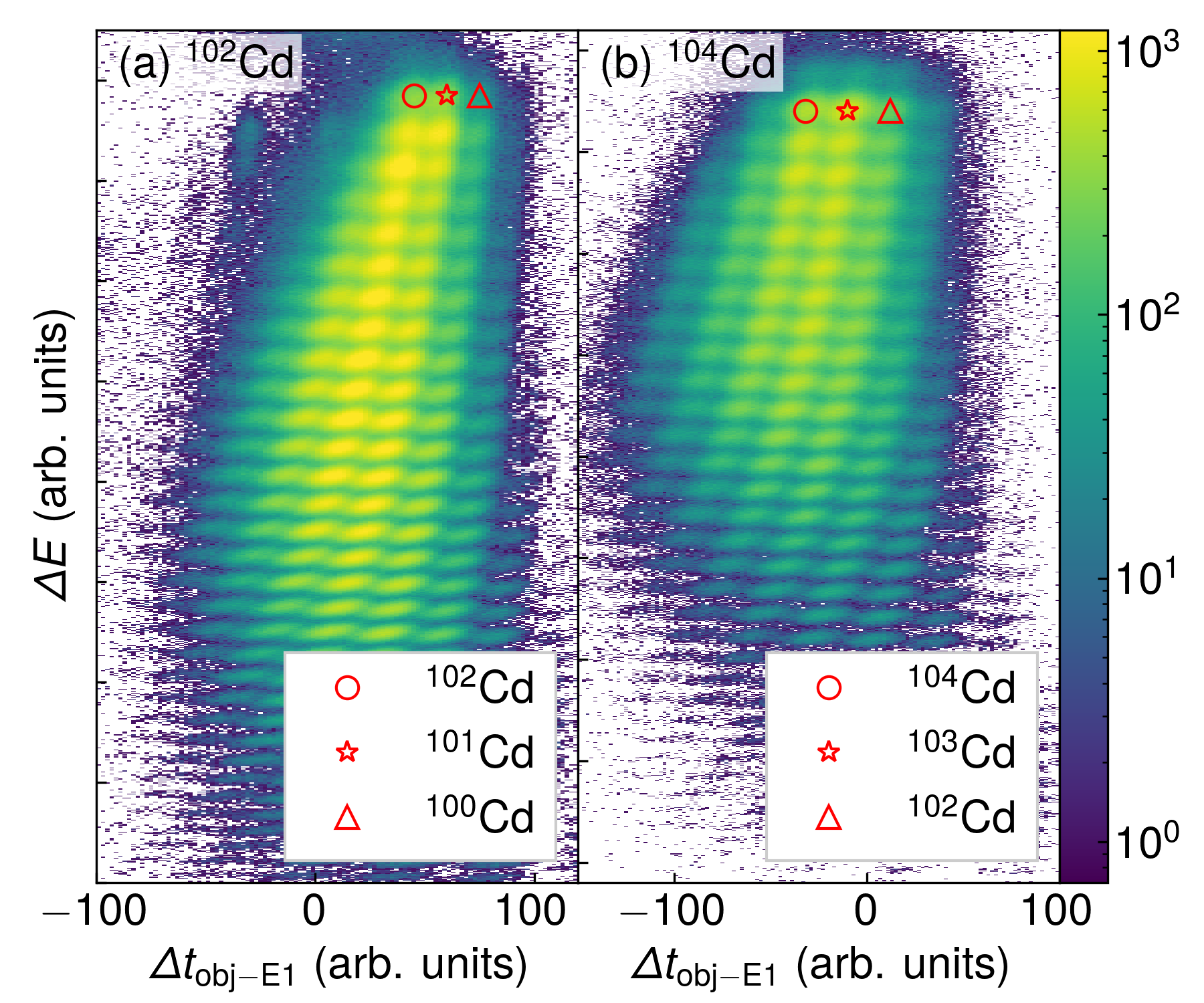}
  \caption{$\gamma$-gated residue PID for (a) incoming $^{102}$Cd, and (b) incoming $^{104}$Cd. Residue species of interest are marked: inelastic scattering (circle), as well as 1n (star) and 2n (triangle) removal channels. }
  \label{fig:residue_pid}
\end{figure}

\paragraph*{}
For a given incoming beam species, a variety of reaction channels
are collected at the focal plane of the S800. These can
be separated on an event-by-event basis using the $\Delta E$-TOF
method, where the energy loss through the ionization chamber at the S800
focal plane is compared to the time of flight, after appropriate
corrections have been applied. Residue particle identification (PID)
plots are shown in Figs.~\ref{fig:residue_pid}(a) and (b) for incoming
$^{102,104}$Cd beams, respectively. The residue nuclei of interest for
the present work are indicated with red symbols. Note that these PID
plots are gated on at least one $\gamma$-ray with energy
$E_{\gamma}>100$~keV detected in GRETINA. This suppresses unreacted
beam components.

\begin{figure}[h]
  \includegraphics[width=0.93\columnwidth]{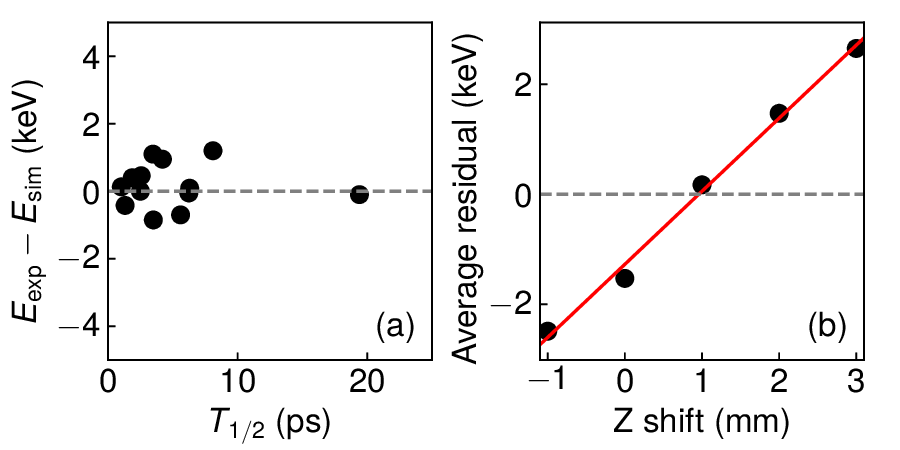}
  \caption{Summary of the $z$-offset optimization. (a) Each point corresponds to a transition with a known lifetime in Cd, Ag, or Pd. The difference between simulated and experimentally observed centroids are shown. (b) Average difference as a function of $z$-offset. The optimzal $z$ offset minimizes this difference and is found to be $1.0$~mm.}
  \label{fig:zpos}
\end{figure}

\newpage
\onecolumngrid
\begin{figure*}[t]
  \includegraphics[width=\linewidth]{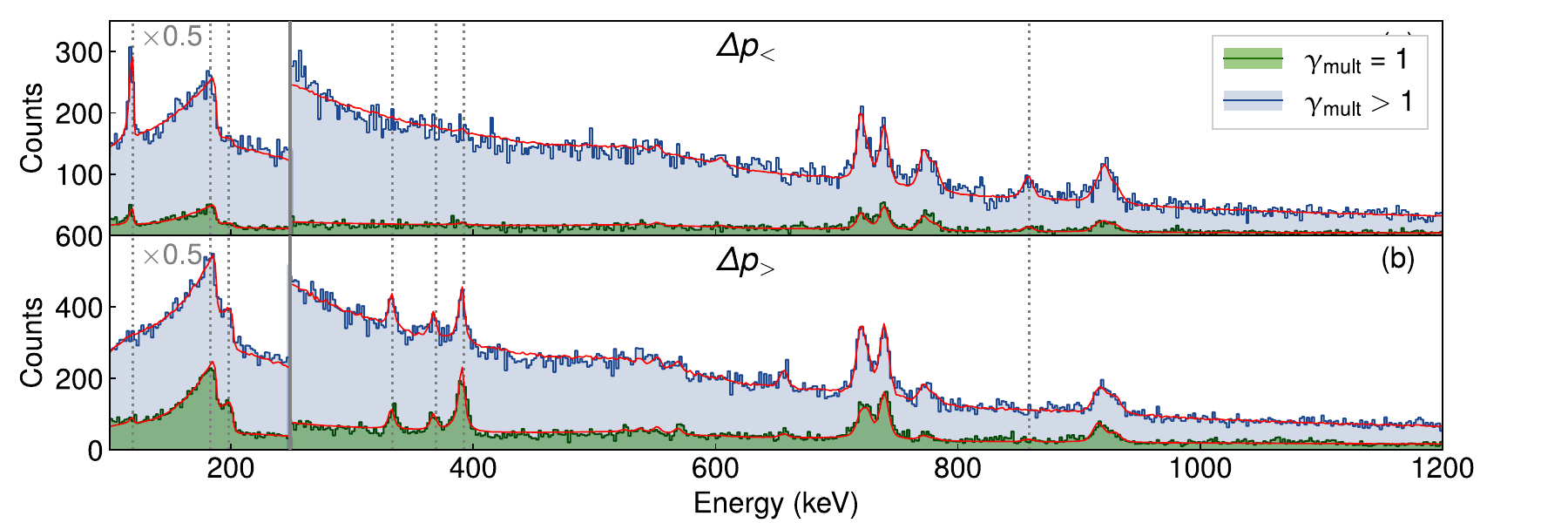}
  \caption{Spectra for $^9$Be($^{104}$Cd,$^{103}$Cd+$\gamma$)$X$
    reaction. $(a)$ is gated on low momenum transfer, i.e. $\Delta
    p_{<}$. $(b)$ is gated on high momentum transfer, $\Delta p_{>}$.
    For both sets of momentum gates, spectra with
    $\gamma_{\mathrm{mult}} = 1$ and $\gamma_{\mathrm{mult}} > 1$ are
    plotted separately. Note the different transitions which are
    enhanced or supressed as a result of the different gate
    combinations. Vertical dashed grey lines mark transitions that
    show strong selectivity to momentum transfer, $\gamma$
    multiplicity, or both. The lowest part of the spectra have been
    scaled by $0.5$ for display purposes.} %
  \label{fig:cd103_spectra}
\end{figure*}
\twocolumngrid

The average effective beam spot angle and position were adjusted to
optimize $\gamma$-ray resolution in the Doppler-corrected spectra. In
addition, the effective $z$-offset of the target was determined by
fits to transitions from multiple states with known lifetimes. For
each state, the peak of the simulated response was compared to the
experimentally observed decay response (Fig.~\ref{fig:zpos}(a)). The differences were plotted as a function of $z$-offset, as shown in Fig.~\ref{fig:zpos}(b). This fixes the $z$ offset at $\approx 1.0$~mm downstream of the nominal target position. Note that lifetime-dependent effects are fully incorporated by the simulation; this is also demonstrated in the lack of correlation present in Fig.~\ref{fig:zpos}(a).

\subsection{Fitting}
The $\gamma$-ray spectra were fitted using the UCGretina GEANT4
framework~\cite{Riley2021,GEANT4,Tange2025}. A response function for each state
populated in the residual nucleus was simulated. The fit parameters
were amplitudes corresponding to direct population of each excited state. 

\paragraph*{}
Nucleon-removal reactions around \mbox{$80-100$~$A$MeV} and above often
involve complex background shapes. A major component of the observed
background is induced by neutrons which are produced at the target ---
either directly incident on the detectors, or interacting with nearby
material to produce $\gamma$ rays. To account for this,
forward-focused neutron distributions were simulated using UCGretina.
These resulted in a variety of continuous backgrounds, as well as
un-Doppler shifted (stopped) peaks from neutron scattering. To most
accurately account for these backgrounds, both the Doppler-corrected
and the lab-frame $\gamma$-ray spectra were fit simultaneously with
simulated Doppler-corrected and lab-frame response functions. The
lab-frame spectra constrain the neutron-induced backgrounds to a large
extent as a result of the stopped transitions. Conversely, the
Doppler-corrected spectra constrain the amplitudes of the states
populated in the residual nucleus. In addition to the simulated
neutron-induced backgrounds, a smooth exponential background was also
used.

\section{Results}

\subsection{$^{103}$Cd}

$\gamma$-ray spectra for the $^9$Be($^{104}$Cd,$^{103}$Cd+$\gamma$)$X$ reaction are shown in Fig.~\ref{fig:cd103_spectra}. The total $\gamma$-ray spectrum is split into four sub-spectra, relating to the momenum transfered $\Delta p$ and the number of detected $\gamma$ rays, $\gamma_{\mathrm{mult}}$. Both of these quantities are useful for discriminating the population of different kinds of states. The low-momenum cut $\Delta p_{<}$ preferrentially populates states with higher spin and more complex structure. This provides some selectivity for states populated by more central collisions, resulting in dissipative reactions~\cite{Gade2022,Gade2022b}. The high multiplicity $\gamma_{\mathrm{mult}}$ selects for high-lying states which decay via cascades of many sequential $\gamma$ rays. This additionally has the effect of suppressing time-correlated background. There is also a correlation between $\Delta p_{<}$ and $\gamma_{\mathrm{mult}}$ cuts: the complex, high-spin states populated in the dissipative reactions often decay by cascades of many $\gamma$ rays. Such effects have been observed several times previously in a variety of different nuclei~\cite{Gade2008, Stroberg2014b, Mutschler2016, Crawford2017, Spieker2019, Gade2022, Gade2022b}.

\begin{figure}[t]
  \includegraphics[width=0.9\columnwidth]{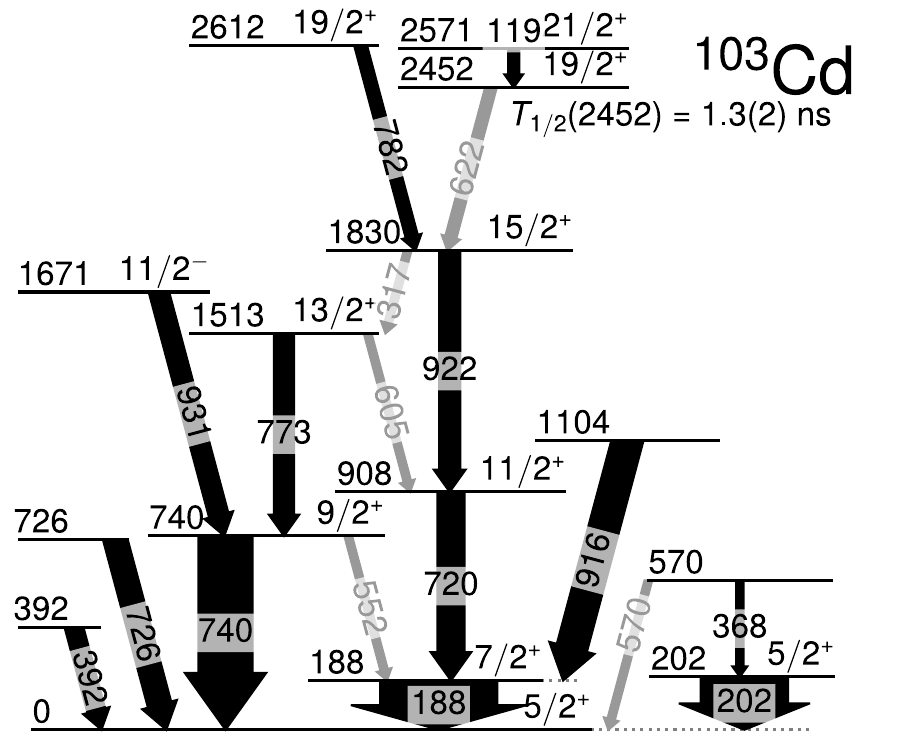}
  \caption{Adopted partial level scheme for $^{103}$Cd. Arrow widths
    are approximately proportional to the observed intensity in the
    $^9$Be($^{104}$Cd,$^{103}$Cd+$\gamma$)$X$ reaction. {Grey
    transitions were not observed directly in this work, but were
    included in the GEANT4 simulations, based on Ref.~\cite{DeFrenne2009}}. All spin
    asisgnments are tentative, based on
    Ref.~\cite{DeFrenne2009}. The 2452-keV state has a long half-life of $T_{1/2} = 1.3(2)$~ns~\cite{Palacz1997}, explaining why the 119-keV peak is strongly present, but no 622-keV peak can be seen. The Doppler-shifted 622-keV transition is nonetheless present in the GEANT4 simulation.} %
  \label{fig:cd103scheme}
\end{figure}

\paragraph*{}
The adopted level scheme for $^{103}$Cd is shown in Fig.~\ref{fig:cd103scheme}.
Several new transitions were observed at energies 334(1)-keV,
658(1)-keV, and 859(2)-keV~\cite{supplementary}. There were insufficient statistics to
place these in the level scheme with $\gamma$-$\gamma$ coincidences,
however we note that the 859-keV was only present in the spectra with
high momentum transfer, while the 658-keV and 334-keV transitions were
only present in the spectra with low momentum transfer. Systematic
uncertainties corresponding to the adopted $\beta = v/c$ and target
$z$-offset positions were included.

\begin{figure}[h]
  \includegraphics[width=\columnwidth]{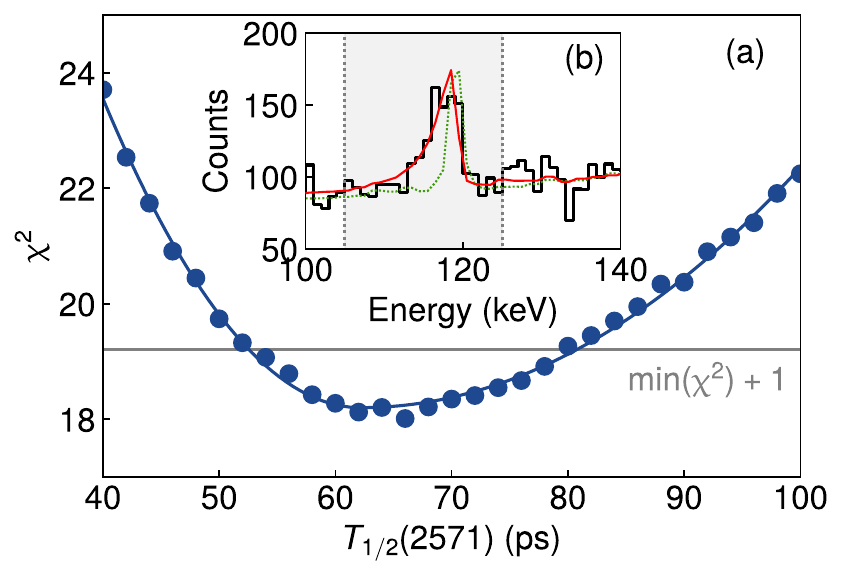}
  \caption{The half-life of the 2571-keV state can be extracted from
    the energy shift in the observed 119-keV transition. (a) shows the
    $\chi^2$ as a function of $T_{1/2}(2571)$, while (b) shows the
    relevant section of the Doppler-corrected energy spectrum. The
    solid red line corresponds to the best-fit half-life of $T_{1/2} =
    67$~ps, while the dotted green line shows the simulated spectrum
    if $T_{1/2} = 0$ is assumed instead. The $\chi^2$ was evaluated in the shaded grey region.} %
  \label{fig:lt2571}
\end{figure}

\paragraph*{}%
Several half-lives could be extracted from lineshape analysis. The 2571-keV state is populated strongly, and decays by a single, 119-keV $\gamma$ ray~\cite{Palacz1997}. However, the peak observed in the knockout reaction is shifted to lower energies by $\approx 2$~keV. The half-life can be extracted as $T_{1/2}(2571) = 67(14)_{\mathrm{stat}}$~ps by plotting the quality of fit as a function of the 2571-keV state half-life, see Fig.~\ref{fig:lt2571}. An additional $10$~ps of systematic uncertainty is added to account for a $\pm 0.5$~mm uncertainty on the effective $z$-position of the target, giving a final value of $T_{1/2}(2571) = 67(14)_{\mathrm{stat}}(10)_{\mathrm{sys}} = 67(17)$~ps.

The half-lives of low-lying 188-keV and 202-keV states can also be estimated. The half-life of the 188-keV state has been measured previously in a LaBr$_3$ fast timing measurement, $T_{1/2}(188) = 370(30)$~ps~\cite{Kisyov2011}. In the present data, the line shape of the 188-keV transition is combined with a broadened 202-keV transtion.

\begin{figure}[b]
  \includegraphics[width=\columnwidth]{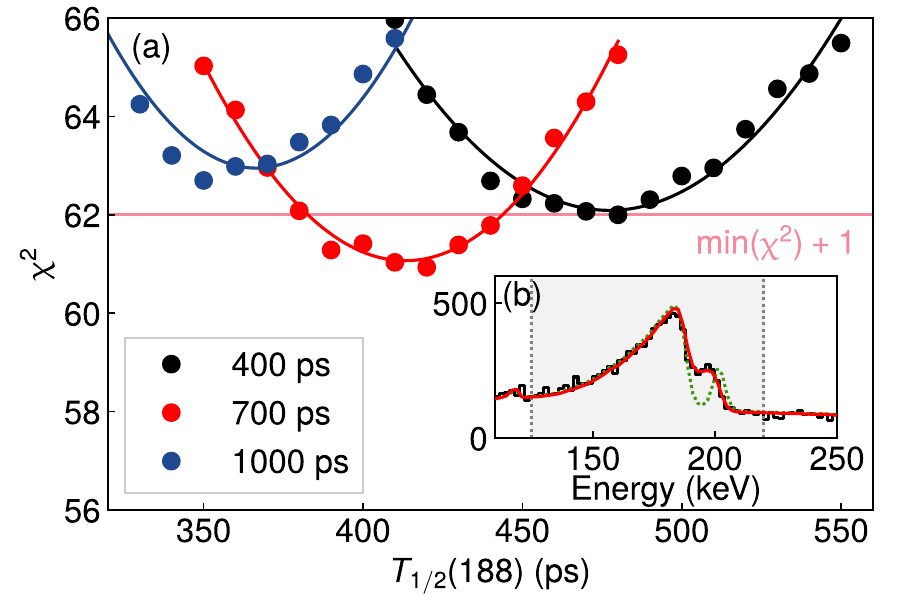}
  \caption{Half-life of the 188-keV state extracted under three
    assumptions for $T_{1/2}(202)$. (a) shows three
    $\chi^2$ curves for $T_{1/2}(202) = 400, 700$, and $1000$~ps. (b)
    shows the necessity of a longer-lived 202-keV state: the dotted
    green curve shows the best fit with $T_{1/2}(202) = 0$~ps, while
    the red shows a representative good fit with $T_{1/2}(202) =
    700$~ps. The $\chi^2$ was evaluated in the shaded grey region.} %
  \label{fig:lt188}
\end{figure}

\paragraph*{}
Half-life $\chi^2$ curves for the 188-keV state were produced under
three assumptions for the 202-keV half-life: 400~ps, 700~ps, and
1000~ps. These represent realistic lower and upper limits based on the
shifted 202-keV peak location in the $\Delta p_{>}$ gated spectra. The
inset panel (b) shows the necessity of a longer half-life for the
202-keV state: the dotted green curve is the best fit with a short half-life
for this state, while the red represents a representative good fit
with $T_{1/2}(202) = 700$~ps. The three curves corresponding to the
different assumptions for $T_{1/2}(202)$ are shown in
Fig.~\ref{fig:lt188}(a), and combined give a half-life estimate of
$T_{1/2}(188) = 410(30)_{\mathrm{stat}}(90)_{\mathrm{sys}}$~ps, where
the systematic uncertainty reflects the influence of the 202-keV
half-life, as well as the shape of the continuous background beneath the
188-keV peak. The data with $\gamma_{\mathrm{mult}} = 1$ was used to
extract the half-life; this reduces sensitivity to the shape of the
low-energy background. The measured half-life is consistent with the
previously reported $T_{1/2} = 370(30)$~ps~\cite{Kisyov2011}, albeit with
much larger uncertainty. 

\subsection{$^{101}$Cd}
\begin{figure}[t]
  \includegraphics[width=\columnwidth]{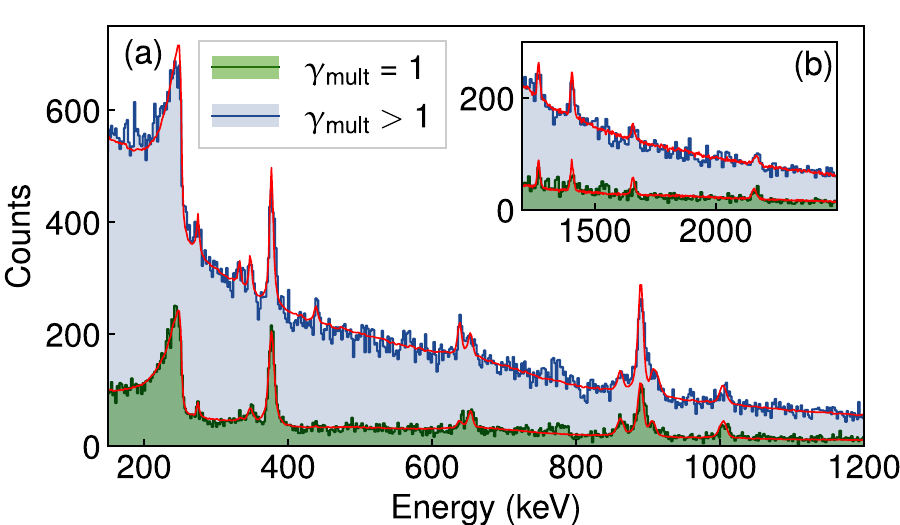}
  \caption{Spectra for the $^9$Be($^{102}$Cd,$^{101}$Cd+$\gamma$)$X$ reaction. Spectra with $\gamma_{\mathrm{mult}} = 1$ and $\gamma_{\mathrm{mult}} > 1$ are plotted separately.} %
  \label{fig:cd101_spectra}
\end{figure}

Spectra for the $^9$Be($^{102}$Cd,$^{101}$Cd+$\gamma$)$X$ reaction are given in Fig.~\ref{fig:cd101_spectra}. No momentum cuts have been applied in this case; the lower statistics and fewer populated states make this less instructive. 

\begin{figure}[b]
  \includegraphics[width=\columnwidth]{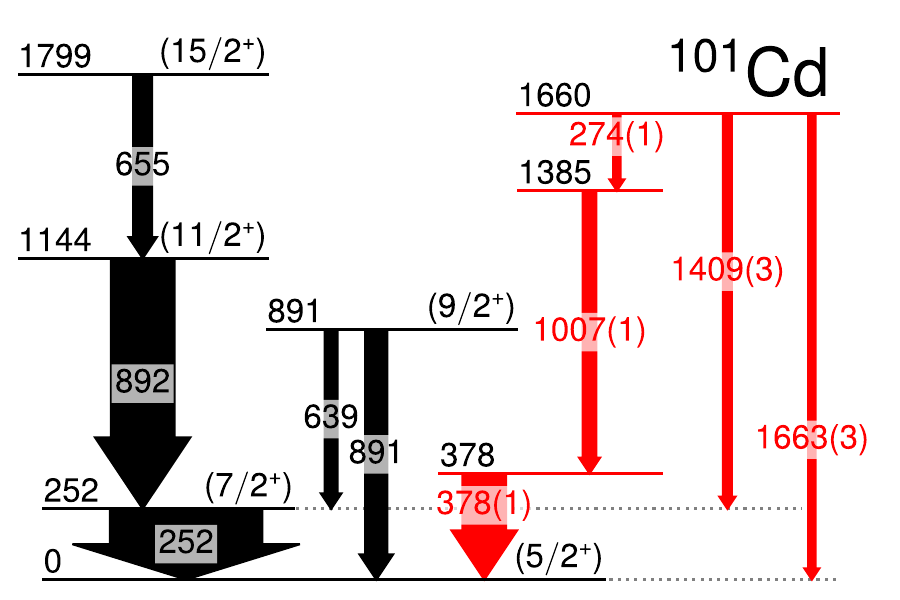}
  \caption{Adopted partial level scheme for $^{101}$Cd. Red
    transitions are newly observed and their placement in the level
    scheme is tentative. Arrow widths are approximately proportional
    to the observed intensity in the
    $^9$Be($^{102}$Cd,$^{101}$Cd+$\gamma$)$X$ reaction. All spin assignments are tentative, based on Ref.~\cite{Blachot2006}. } %
  \label{fig:cd101scheme}
\end{figure}

The adopted level scheme for $^{101}$Cd is shown in
Fig.~\ref{fig:cd101scheme}. Red transitions are newly observed, and
their placement in the level scheme is tentative. In addition to the
transitions shown here, seven transitions with energies of 333(2)~keV,
349(1)~keV, 441(1)~keV, 863(2)~keV, 907(4)~keV, 1270(2)~keV, and
2170(3)~keV were present but could not be placed in the level scheme~\cite{supplementary}.

\begin{figure}[h]
  \includegraphics[width=\columnwidth]{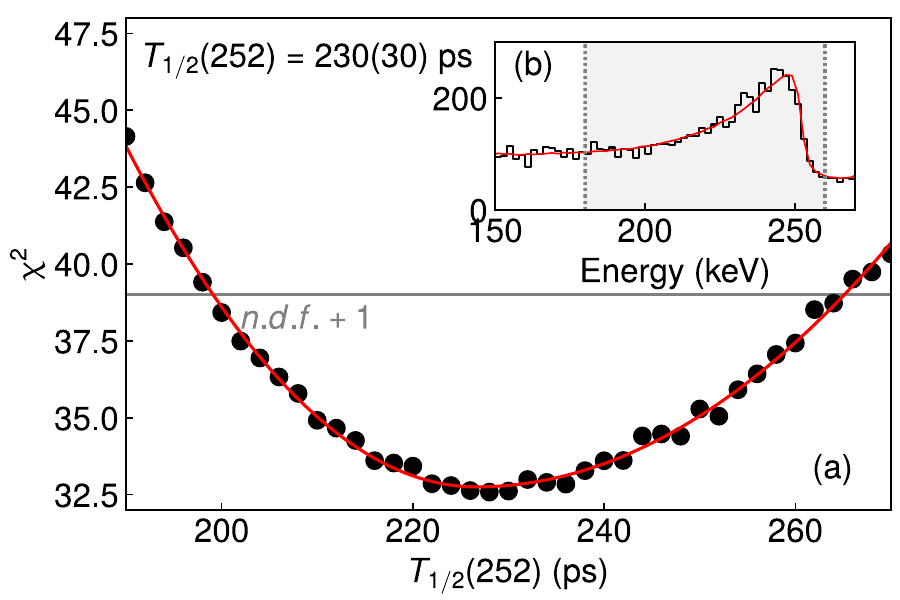}
  \caption{$\chi^2$ curve used to extract the half-life of the 252-keV
    state in $^{101}$Cd. The $\chi^2$ was evaluated in the shaded grey region.} %
  \label{fig:lt252}
\end{figure}

Similarly to the case of $^{103}$Cd, the half-life of the low-lying 252-keV state could be established through its lineshape. The $\chi^2$ as a function of half-life is shown in Fig.~\ref{fig:lt252}, and this establishes a new half-life for this state, $T_{1/2} = 230(30)$~ps. 

\subsection{$^{103}$Sn}
\begin{figure}[b]
  \includegraphics[width=0.7\columnwidth]{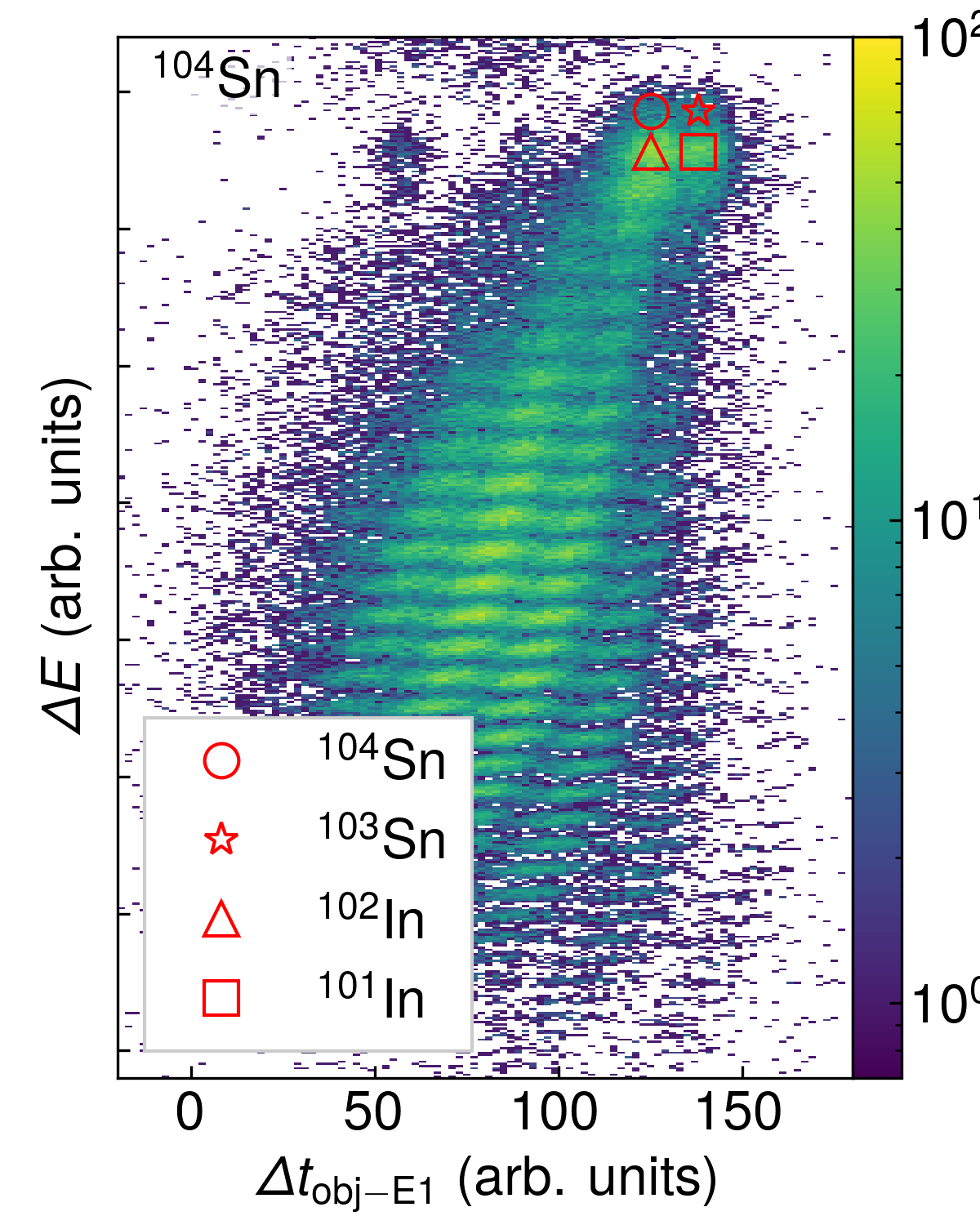}
  \caption{Residue particle identification for incoming $^{104}$Sn
    beam. The primary residues of interest, $^{104,103}$Sn are marked.
    In addition, $^{102,101}$In species are marked; they are strongly
    present due to contamination in the incoming particle identification.} %
  \label{fig:sn104_pid}
\end{figure}

\begin{figure}[h]
  \includegraphics[width=\columnwidth]{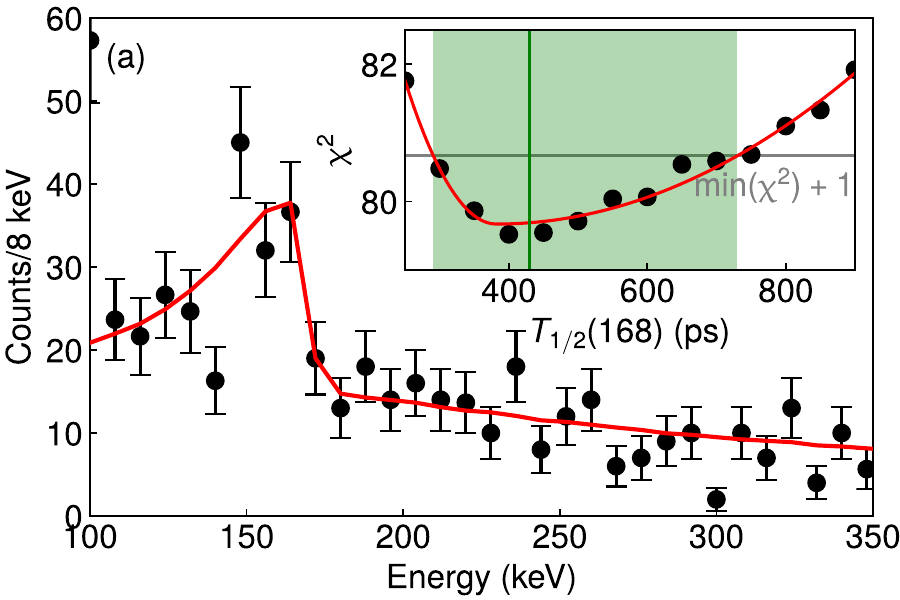}
  \caption{(a) Spectrum from the $^9$Be($^{104}$Sn,$^{103}$Sn+$\gamma$)$X$ reaction. Only a single $\gamma$-ray transition is observed, depopulating the 168-keV state. An estimate of the half-life of the state is extracted in panel (b): $T_{1/2} = 430(^{+300}_{-220})$~ps. } %
  \label{fig:sn103_spectra}
\end{figure}

Despite very low statistics, the $^9$Be($^{104}$Sn,$^{103}$Sn+$\gamma$)$X$ reaction was investigated. The residual particle identification for an incoming $^{104}$Sn beam is shown in Fig.~\ref{fig:sn104_pid}. The resulting $\gamma$-ray spectrum is shown in Fig.~\ref{fig:sn103_spectra}(a), with the inset showing the sensitivity to the half-life of the 168-keV state, measured here as $T_{1/2} = 430(^{+300}_{-220})$~ps.

A summary of the half-lives and corresponding $B(M1)$ strengths are
given in Table~\ref{tab:lt_results}. While the $\delta(M1/E2)$ mixing
ratios are not known, limits of $B(E2; 7/2^+ \rightarrow 5/2^+)\leqslant
5$~W.u. have been set for the purposes of extracting $B(M1)$
strengths, and are included as an additional source of uncertainty.
The effect on the extracted $B(M1)$ is rather small however, since the
decay is dominated by the $M1$ component even with relatively large
$B(E2)$ matrix elements. Conversion coefficients from the BrIcc database~\cite{BrIcc} were used.

\begin{table}
\begin{ruledtabular}
  \caption{Half-lives and corresponding $B(M1)$ strengths
    from this work. $B(M1)$ strengths are extracted under the
    assumption that $B(E2; 7/2^+ \rightarrow 5/2^+) \leqslant 5$~W.u.,
    the effect of the $E2$ strength on the half-life is very small.}
  \label{tab:lt_results}
  \begin{tabular}{cSSr}
    {Isotope} & {Energy (keV)} & {$T_{1/2}$ (ps)} & {$B(M1)$
      ($\mu_N^2$)} \\ \hline
    {$^{101}$Cd} & 252 & 230(30) & 0.0103(17) \\
    {$^{103}$Cd} & 188 & {410(95)} & 0.0134(33) \\
    {$^{103}$Sn} & 168 & {$430(^{+300}_{-220})$} & $0.0171(^{+180}_{-70})$ \\
  \end{tabular}
\end{ruledtabular}
\end{table}

\section{Discussion}
To assist with the interpretation of the measured half-lives, shell
model calculations for neutron-deficient Cd and Sn nuclei were conducted.
The ``jj45b'' interaction was used. The ``jj45b'' Hamiltonian is based on the two-body matrix elements
(TBME) derived in Ref.~\cite{Shergur2002} for the $^{132}$Sn region, with the proton-proton
$T=1$ TBME replaced by those for $N=50$ obtained in Ref.~\cite{Lisetskiy2004}, and neutron
single-particle energies adjusted to reproduce spectra of $N=51$ nuclei near $^{100}$Sn.
The model space had active $\pi 0f_{5/2},
1p_{3/2}, 1p_{1/2}$, and $0g_{9/2}$ proton orbitals, and active $\nu
1d_{5/2}, 0g_{7/2}, 1d_{3/2}, 2s_{1/2}, 0h_{11/2}$ neutron orbitals.
The calculations were performed using the ``KSHELL''
program~\cite{KSHELL}. 

\begin{figure}[h]
  \includegraphics[width=\columnwidth]{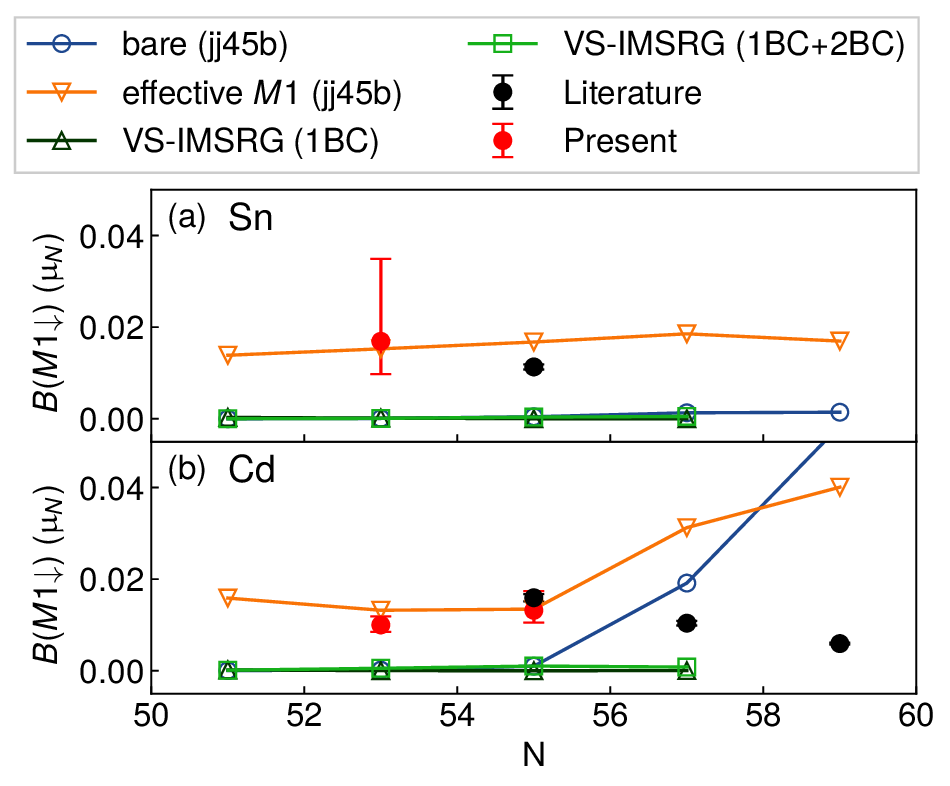}
  \caption{Systematics of $B(M1; 7/2^+ \rightarrow 5/2^+)$ transitions
  in neutron-deficient Sn and Cd isotopes. Literature half-lives are from
  Refs.~\cite{Kisyov2011, Pasqualato2023, Blachot2008,Lalkovski2019}.} %
  \label{fig:bm1_syst}
\end{figure}

The $B(M1;7/2^+ \rightarrow 5/2^+)$ strengths for the region are shown
in Fig.~\ref{fig:bm1_syst}. Experimentally, these strengths are small
but non-zero. The ``jj45b'' calculations indicate that the $7/2^+ \rightarrow
5/2^+$ transition corresponds to a $0g_{7/2}\rightarrow 1d_{5/2}$
$l$-forbidden $M1$ transition. Using the bare $M1$ operator, these
transitions are vanishingly small, however, the effective operator given by
Eq.~\ref{eq:effective_M1} is often used to produce more realistic results. Notably, the orbital and spin
terms can be empricially constrained by the wealth of data available
for ground-state magnetic moments and $B(GT)$ strengths. However,
$l$-forbidden $M1$ transitions, which are the predominant constraint
on the tensor term $g_P$ are reasonably rare --- especially near
double shell closures.

The bare $M1$ operator gives vanishing $B(M1)$ strength for $N \leq
55$ (see Fig.~\ref{fig:bm1_syst}). If the effective $M1$
operator is used (i.e., Eq.~\ref{eq:effective_M1}) with
bare values for $g_L$, $g_S = 0.7 \times g_{s_{\mathrm{free}}}$, and
$g_P^n = -2.3$, then the experimental $M1$ strengths are reproduced
reasonably well for $N<56$ (see orange points in
Fig.~\ref{fig:bm1_syst}). Microscopic calculations for the corrections
to the $M1$ operator for the region of $^{132}$Sn are given in Table VI
of Ref.~\cite{Brown2005} where a value of $g_P^n = -1.2$ was obtained for
the $0g_{7/2}$ to $1d_{5/2}$ neutron transition. These calculations for $^{132}$Sn
explicitly included core-polarization,
mesonic-exchange currents, $\Delta$-isobar admixtures and
relativistic corrections. Similar calculations
for the $^{100}$Sn and other regions need to be carried out.

In addition to the ``jj45b'' interaction,
calculations using
VS-IMSRG~\cite{Stroberg2019,Miyagi2022,Stroberg2017} using the 1.8/2.0
(EM) interaction~\cite{Hebeler2011,Entem2003} with the imsrg++
code~\cite{imsrg} were run. The valence space was the same as the
``jj45b'' calculations. $M1$ strengths were extracted from the
VS-IMSRG calculations using a microscopically-derived effective
operator, initially with only one-body
currents considered (1BC), but also with the effects of one- and two-body
currents (1BC+2BC). See Ref.~\cite{Miyagi2024} for a discussion of
two-body (or meson-exchange) currents in the VS-IMSRG. In both cases,
the calculated $M1$ strengths for the $7/2^+ \rightarrow 5/2^+$ transition are far below the experimental values,
effectively the sames as those extracted with the bare $M1$ operator.

\begin{figure}[t]
  \includegraphics[width=\columnwidth]{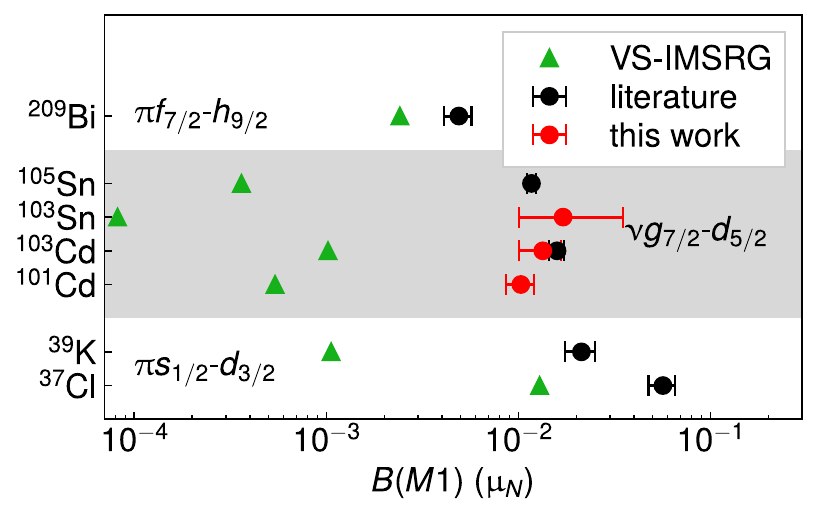}
  \caption{Comparison of $l$-forbidden $M1$ strengths between VS-IMSRG
  and experiment for the $^{100}$Sn region, as well as several nuclei near
  double shell closures reported in Ref.~\cite{Miyagi2024}. The VS-IMSRG underpredicts the $M1$ strength across
  the chart of the nuclides. Literature $M1$ strengths are extracted
  from Refs.~\cite{Pasqualato2023, Cameron2012,Chen2015,Chen2018}.} %
  \label{fig:vsimsrg_global}
\end{figure}

This disagreement is in contrast to the generally good agreement found
for VS-IMSRG calculations with ground-state magnetic moments across
the nuclear chart in Ref.~\cite{Miyagi2024}. To investigate this
further, $l$-forbidden $M1$ transitions for $^{37}$Cl, $^{39}$K, and
$^{209}$Bi were also calculated with VS-IMSRG. The results are shown
in Fig.~\ref{fig:vsimsrg_global}. These three nuclei are
near double shell closures, and all see improvements to the magnetic
moment agreement with the inclusion of two-body currents~\cite{Miyagi2024}. Despite this, VS-IMSRG
seems to globally underpredict $l$-forbidden $M1$ strengths. The best
agreement is found for $^{209}$Bi, though the theoretical $M1$ strength is still roughly a
factor of 2 too small.

It is clear that predictions of $l$-forbidden $M1$ strengths remain a
challenge for \textit{ab initio} theory, and these results motivate
future theoretical work to better describe the effective $M1$ operator
from first principles. In particular, the effects of the $\Delta(1232)$ isobar state are not presently included in the VS-IMSRG formulation of the effective $M1$ operator. The inclusion of the $\Delta(1232)$ state may enable more accuate predictions of nuclear $M1$ properties from \textit{ab initio} calculations~\cite{Towner1983,Knpfer1983}.

\section{Conclusion}
Knockout reactions on $^{104,102}$Cd and $^{104}$Sn beams have been performed at the
Facility for Rare Isotope beams using GRETINA and the S800. A variety
of states are populated, both ``simple'' quasiparticle states as well
as more complex structures. The level schemes for $^{103,101}$Cd have
been extended with 3 and 12 new $\gamma$ rays, respectively. Several
new half-lives have been measured using the Doppler broadened
lineshapes observed in GRETINA. These half-lives can be related to the
$B(M1; 7/2^+ \rightarrow 5/2^+)$ strength, which is small but
non-zero.

The $l$-forbidden $B(M1)$ strengths provide a constraint on
the $g_P$ coupling term of the effective $M1$ operator for the
$^{100}$Sn region. They have been compared to the results of VS-IMSRG
calculations, with two-body currents included. Despite this, poor
agreement with the VS-IMSRG results is found; it seems that the
VS-IMSRG underpredicts $l$-forbiddent $M1$ strengths globally. It is
unclear whether this is the result of model spaces that are too small,
or if further refinements to the VS-IMSRG methods are needed. 

\begin{acknowledgements}
Thanks to Ragnar Stroberg and Thomas Papenbrock for helpful discussions. This material is based upon work supported in part by the U.S. Department of Energy, Office of Science, Office of Nuclear Physics under Contract Nos. DE-SC0023633 (Michigan State), DE-AC05-00OR22725 (ORNL), DE-FG02-96ER40983 (UTK), and DE-AC02-05CH11231 (LBNL). The publisher acknowledges the US government license to provide public access under the DOE Public Access Plan (http://energy.gov/downloads/doe-public-access-plan). This work was also supported by the U.S. National Science Foundation under Grant No.~PHY-2110365 (MSU), and through the FRHTP program under award No. PHY-2402275. This material is based upon work supported by the U.S. Department of Energy, Office of Science, Office of Nuclear Physics and used resources of the Facility for Rare Isotope Beams (FRIB) Operations, which is a DOE Office of Science User Facility under Award Number DE-SC0023633. This research was also supported in part by Notre Dame’s Center for Research Computing.
\end{acknowledgements}

\bibliography{e21003_abbrev}

\end{document}


\section{Supplemental Material}

Tables~\ref{tab:state_populations} and \ref{tab:rel_intensities} give
the relative population amplitudes and $\gamma$-ray intensities
following the $^9$Be($^{104}$Cd,$^{103}$Sn+$\gamma$)$X$ and
$^9$Be($^{102}$Cd,$^{101}$Cd+$\gamma$)$X$ reactions. 

\paragraph*{}
Uncertainties for
newly-observed transition energies are also given in
Table~\ref{tab:rel_intensities}. For previously observed transitions,
the energies obtained from Refs.~\cite{DeFrenne2009,Blachot2006} are
far more precise than the present experiment due to the strong Doppler
broadening. In the present work, sometimes only partially resolved doublets and triplets
are observed. As such, the present measurement should not be taken as
a measurement of these transition energies, but the energies are given
only as a label.

\vspace{7ex}

\begin{table}[h]
\begin{ruledtabular}
  \caption{State population amplitudes obtained from the present work, normalized to the strongest-populated state. }
  \label{tab:state_populations}
  \begin{tabular}{SSSS}
    \multicolumn{2}{c}{$^{103}$Cd}             & \multicolumn{2}{c}{$^{101}$Cd} \\ \cmidrule{1-2} \cmidrule{3-4}
    {Energy (keV)} & {Amplitude} & {Energy (keV)} & {Amplitude} \\
    188            &100(8)       &   252          &17.6(11) \\
    202            &92(6)        &   891          &57(10)    \\   
    392            &26.9(21)     &   1144         & 100(9)   \\ 
    570            &13.7(32)     &  1799          & 31(4)    \\ 
    726            &35(4)        &   333          & 8.5(25)    \\ 
    740            &36(4)        &   349          &17.0(27)  \\   
    908            &33(4)        &   378          &64(5)     \\   
    1104           & 31.3(34)    &   441          &8.2(27) \\     
    1513           & 31.7(28)    &   863          &31(4)  \\ %
    1671           & 18.1(26)    &   907          &47(4)     \\ 
    2571           & 10.8(14)    & 1270           & 26(4)  \\
    2612           & 20.6(22)    & 1385           &33(4)      \\ 
    334            &10.5(18)     & 1660           &66(6)    \\   
    658            &13.8(22)     & 2170           & 16(4)    \\    
    859            &15.7(25)     &                & \\
  \end{tabular}
\end{ruledtabular}
\end{table}

\newpage

\begin{table}[t]
\begin{ruledtabular}
  \caption{Relative intensities for $\gamma$-ray transitions following
    $^9$Be($^{104}$Cd,$^{103}$Sn+$\gamma$)$X$ and
    $^9$Be($^{102}$Cd,$^{101}$Cd+$\gamma$)$X$ reactions. These are
    extracted from the amplitudes in Table~\ref{tab:state_populations}
    and branching ratios from Refs.~\cite{DeFrenne2009,Blachot2006} where known, and
    the present data for new states. Uncertainties are statistical
    only from the present work.}
  \label{tab:rel_intensities}
  \begin{tabular}{cScS}
    \multicolumn{2}{c}{$^{103}$Cd}             & \multicolumn{2}{c}{$^{101}$Cd} \\ \cmidrule{1-2} \cmidrule{3-4}
    {Energy (keV)} & {$I_{\gamma}$} & {Energy (keV)} & {$I_{\gamma}$} \\
119                &  5.3(7)        &    252         &  100(6)   \\           
188                &  100(5)        &    274(1)      &  6.3(14)  \\     
202                &  51.0(31)      &    333(2)      &  5.1(15)  \\     
334(1)             &  5.1(9)        &    349(1)      &  10.2(16) \\     
368                &  4.3(9)        &    378(1)      &  58(4)    \\      
392                &  13.6(10)      &    441(1)      &  4.9(16)  \\     
658(1)             &  6.7(11)       &    639         &  13.5(21) \\     
720                &  31.9(24)      &    655         &  17.4(22) \\     
726                &  15.5(17)      &    863(2)      &  18.6(22) \\     
740                &  36.3(25)      &    891/892\footnote{Unresolved doublet}         & 100(6)  \\
773                &  14.3(12)      &    907(4)      &  28.0(27) \\     
782                &  10.1(11)      &    1007(1)     &  20.1(27) \\    
859(2)             &  7.6(12)       &    1270(2)     &  15.8(22) \\    
916                &  14.5(16)      &    1409(3)     &  23.0(25) \\    
922                &  15.1(12)      &    1663(3)     &  10.7(23) \\    
931                &  8.8(13)       &    2170(3)     &  9.6(22)               
  \end{tabular}
\end{ruledtabular}
\end{table}


\bibliography{e21003_abbrev}